\documentclass[oldversion]{aa}
\usepackage{graphicx}

\newcommand{\oii}{[OII]$\lambda3727$}
\newcommand{\hb}{H$\beta$}
\newcommand{\oiii}{[OIII]$\lambda5007$}
\newcommand{\ha}{H$\alpha$}

\newcommand{\rr}{$r^{\prime}$}
\newcommand{\mic}{$\mu$m\ }
\begin{document}
\title{
A galaxy overdensity at z=0.401 associated with 
an X-ray emitting structure of warm-hot intergalactic medium
\thanks{
Based on observations made with the Italian Telescopio Nazionale Galileo 
(TNG), operated on the island of La Palma by the Fundaci\'on Galileo Galilei 
of the INAF (Istituto Nazionale di Astrofisica), and with
the William Hershel Telescope (WHT), operated by the ING, both
at the Spanish Observatorio del Roque de los Muchachos 
of the Instituto de Astrofisica de Canarias.
}\\
}

\author{
       F. Mannucci,  \inst{1}
  \and G. Bonnoli,   \inst{2}
  \and L. Zappacosta \inst{3}
  \and R. Maiolino,  \inst{4}
  \and M. Pedani     \inst{5}
}

\offprints{F. Mannucci \email{filippo@arcetri.astro.it}}

\institute{
     INAF, Istituto di Radioastronomia, 
	 Largo E. Fermi 5, 50125 Firenze, Italy
\and Universit\`a di Siena, Dipartimento di Fisica, via Roma 56, 
     53100, Siena, Italia
\and Department of Physics and Astronomy, 4129 Frederick Reines Hall, 
     University of California, Irvine, CA 92697-4575.
\and INAF, Osservatorio Astronomico di Roma, 
     via di Frascati 33, 00040, Monteporzio Catone (RM), Italia
\and Fundaci\'on Galileo Galilei and Telescopio Nazionale Galileo, 
  P.O. Box 565, E-38700 Santa Cruz de La Palma, Tenerife, Spain
}

\authorrunning{F. Mannucci et al.}
\titlerunning{A galaxy overdensity at z=0.4 associated with the WHIM}

\date{Received / Accepted}

\abstract{
We present the results of spectroscopic observations of 
galaxies associated with the diffuse X-ray emitting
structure discovered by Zappacosta et al. (2002). 
After measuring the redshifts of 161 galaxies,
we confirm an overdensity of galaxies
with projected dimensions of at least 2 Mpc,
determine its spectroscopic redshift in z=0.401$\pm$0.002, 
and show that it is spatially coincident with the
diffuse X-ray emission.
This confirms the original claim that this X-ray emission
has an extragalactic nature and is due to
the warm-hot intergalactic medium (WHIM).
We used this value of the redshift to compute the
temperature of the emitting gas. The resulting value depends on the 
metallicity that is assumed for the IGM, and is constrained to be
between 0.3 and 0.6 keV for metallicities between 0.05 and 0.3 solar, 
in good agreement with the expectations from the WHIM.

}

\maketitle


\section{Introduction}
\label{sec:intro}

Recent cosmological models (Cen \& Ostriker, 1999; Dav\'e et al. 2001;
Mittaz et al. 2004; Kang et al. 2005)
have predicted the existence of large amounts of
baryonic matter in the local universe dispersed in the intergalactic medium
(IGM). This material, comprising about 1/3 of the total baryonic matter, is
expected to be in the form of tenuous shock-heated warm-hot ($10^5-10^7$ K) 
filaments of gas in moderate overdensities ($\delta\sim10-100$)
tracing the dark matter cosmic web (Dolag et al. 2006), 
the so-called warm-hot intergalactic medium (WHIM).
The existence of these structures is believed to be the 
solution of the ``missing baryons problem'' (Fukugita, Hogan \& Peebles
1998; Fukugita \& Peebles 2004), i.e., that finding that many
more baryons are observed at high redshift than in the local universe.
This is ascribed to the fact that about one third of the ordinary matter 
in the local
universe is yet to be detected, as the density and temperature of the gas in
the large-scale filaments make their detection very difficult.

Several effects contribute to making the detection of the X-ray emission
from this material quite difficult. 
The soft X-ray background is dominated by the local hot bubble and the 
emission of the halo of the Milky Way (see, for example, Bregman \&
Lloyd-Davies 2006). Both these two contributions are 
poorly known and it is difficult to disentangled any diffuse 
extragalactic emission from these bright foreground sources.

\begin{figure*}     
\centering
\includegraphics[width=13cm]{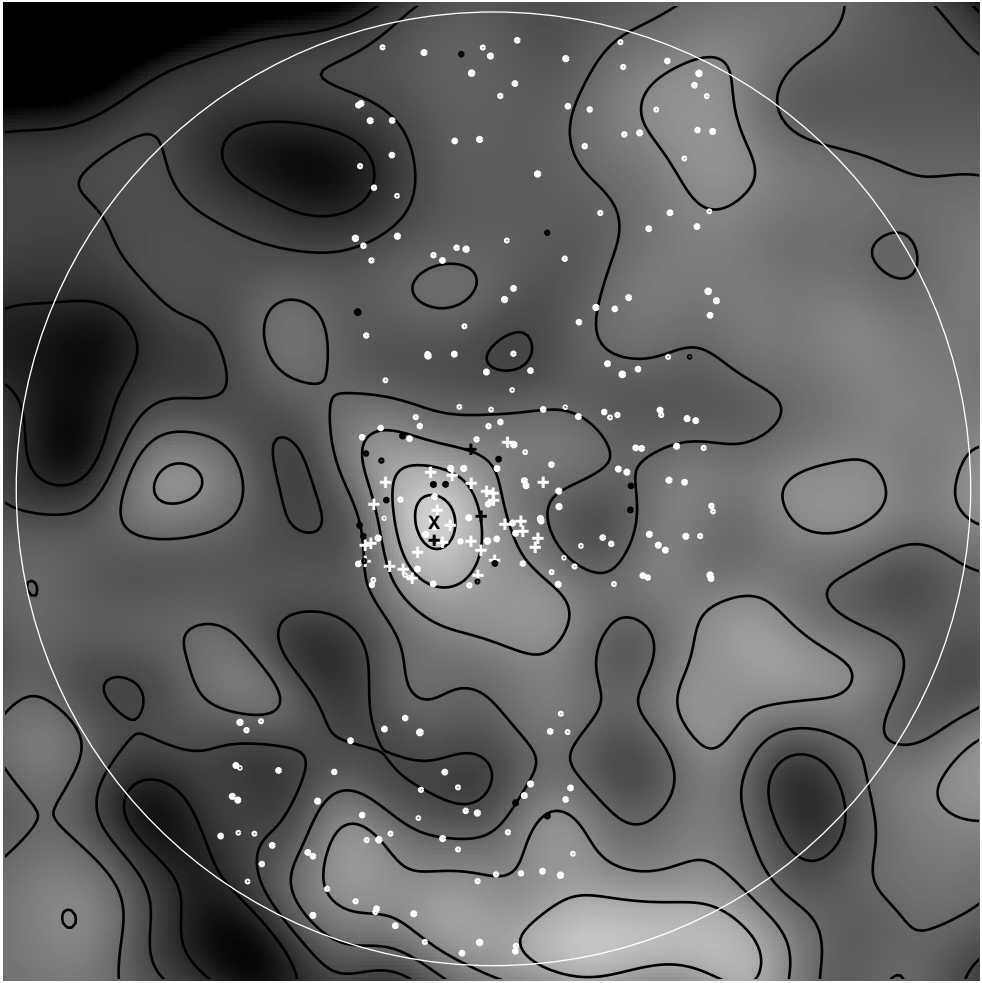}
\caption{
\label{fig:targets}
Distribution of the targets observed with AF2/WYFFOS (dots) 
and DOLORES (crosses)
over the field. The objects with measured redshift between 0.39 and
0.41 and quality class 1 or 2 (see Sect~\ref{sec:redshift}) 
are shown in black, all the other targets in white.
The underlying greyscale image with marked contour levels
is the X-ray brightness 
distribution  in the soft ROSAT/PSPC band described 
in Zappacosta et al. (2002),
with the position of the X-ray peak marked by a black X.
The field is centered at RA=10:10:06.5 DEC=+51:59:56.4 (J2000).
The large white circle has a diameter of 1 deg. 
}
\end{figure*}

Despite these difficulties, several possible detections of
this WHIM have been
reported by using  different techniques
(Zappacosta et al. 2002, 2005a; 
Mathur et al. 2003; 
Soltan et al. 2005; Soltan 2006).
Recently, Dietrich et al. (2005) have demonstrated the correspondence of a
dark matter filament, detected by a weak-lensing study, with a diffuse emission
in the soft X-ray between two clusters of galaxies.

The only detections of WHIM beyond the local universe (z$>$0.1) have been
obtained by imaging in the soft X-ray (kT$<$2keV). 
In particular, in Zappacosta et al. 
(2002) we detected a diffuse soft X-ray structure in a high Galactic latitude 
ROSAT field. Its properties, such as its dimensions (about 7 Mpc),
surface brightness, temperature, and gas density,
are consistent with the expectations for the WHIM according to the predictions
of several models
(Croft et al. 2001; Dolag et al. 2006; Roncarelli et al. 2006;
Ursino \& Galeazzi 2006).
In particular, this structure has a spectral energy distribution (SED)
consistent with plasma
at kT$\sim$0.3 keV, which is too soft to be ascribed to
unresolved AGNs or to clusters of galaxies and is consistent with the
spectrum expected for WHIM. 
Moreover, Zappacosta et al. (2002) demonstrated 
that an overdensity of galaxies is associated with the X-ray structure,
as expected by the models by Viel et al. (2005) and 
Hern\'andez-Monteagudo et al. (2006).
The lack of spectroscopic observations of the galaxies resulted in a 
poor assessment that the overdensity is really present in the 
physical space and is not just a projection effect, as well as in a
very uncertain measure of the distance of this structure.
In fact, the photometric redshift of the galaxies in the overdensity,
z=0.45$\pm$0.15, is highly uncertain and is adequate only
for putting the WHIM far from the local universe.

Since this is the WHIM detection currently at the largest distance, it 
was used by Zappacosta et al. (2005b) to constrain the thermal
history of the WHIM. The temperature of the WHIM is found to 
rapidly decrease with redshift, as expected by several models
(Cen \& Ostriker 1999; Dav\'e et al. 2001).  Confirming the existence 
of such a WHIM and measuring its properties is
therefore very important for constraining the cosmological models.
Measuring the metallicity of the gas and studying the 
interplay between IGM and galaxies would be 
particularly important, as the current cosmological models 
cannot constrain these two aspects.

With this aim, we observed more than 250 galaxies near 
the X-ray peak with two multi-object
spectrographs to study the galaxy overdensity in greater detail, 
confirm its existence, measure its redshift, and 
study the properties of the galaxies.
In the next sections we describe the observations (Sect.~\ref{sec:obs})
and the redshift determination (Sect.~\ref{sec:redshift}). The redshift
distribution, the spatial association with the X-ray peak,
and the properties of the galaxies are discussed in 
Sects.~\ref{sec:zhist} and \ref{sec:spatial}.

Throughout the paper we use the concordance cosmology 
($h_{100},\Omega_m,\Omega_\Lambda$)=(0.7, 0.3, 0.7).

\begin{table}	
\caption{Observation summary}
\begin{tabular}{cccccc}
\hline
\hline
Instrument  & Mag limits &  Number of  & Exp. time \\
            &   (\rr)    &  sources    &  (min)    \\
\hline
WHT/WYFFOS  & 20.0--20.7 &    77       &    60     \\
WHT/WYFFOS  & 20.7--21.1 &    79       &   100     \\
WHT/WYFFOS  & 21.1--21.5 &    81       &   230     \\
TNG/DOLORES & 20.6--21.6 &    32       &   150     \\
\hline
\end{tabular}
\label{tab:obs}
\end{table}
\begin{figure}     
\centering
\includegraphics[width=6.1cm,angle=-90]{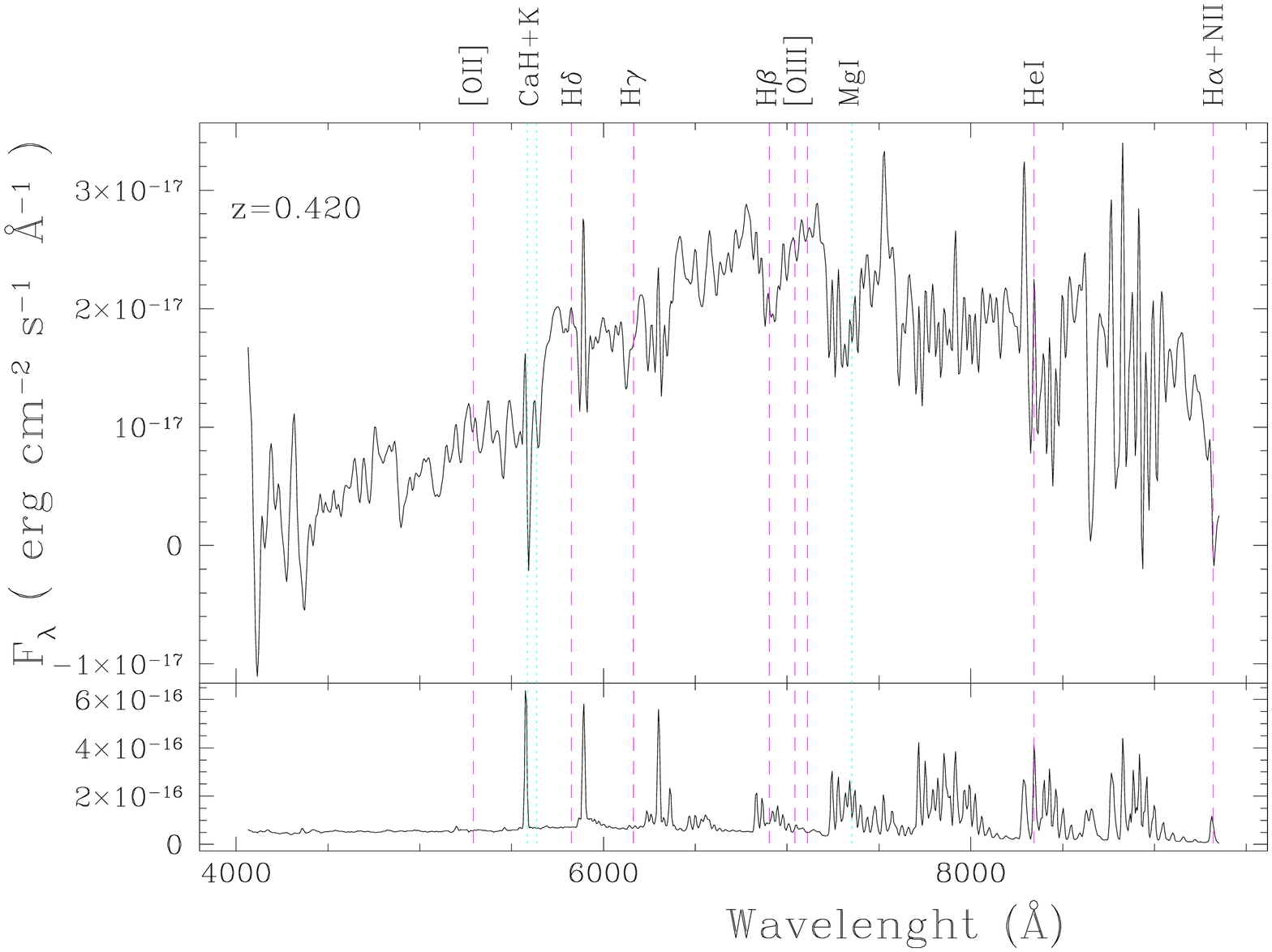}
\includegraphics[width=6.1cm,angle=-90]{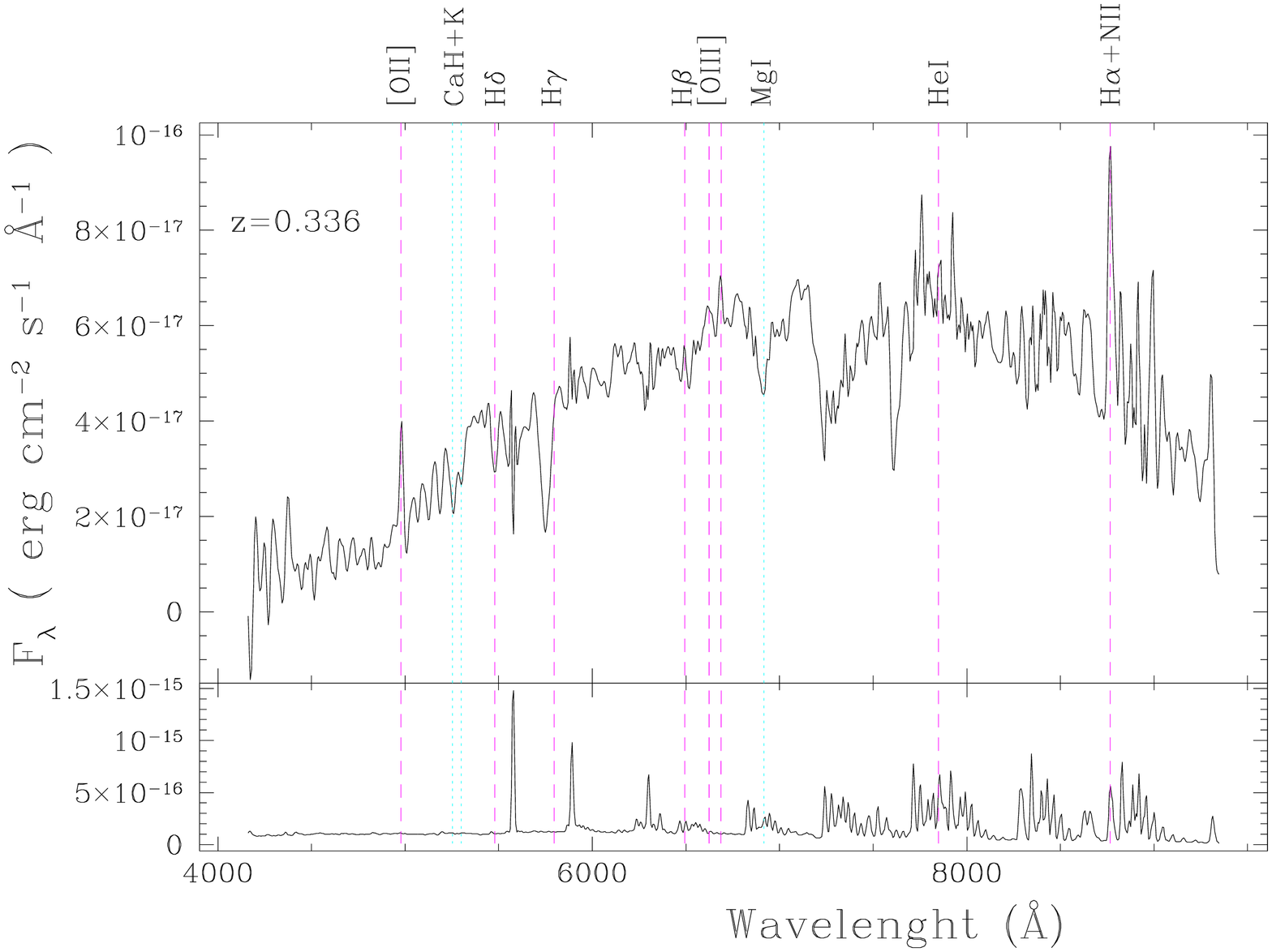}
\includegraphics[width=6.1cm,angle=-90]{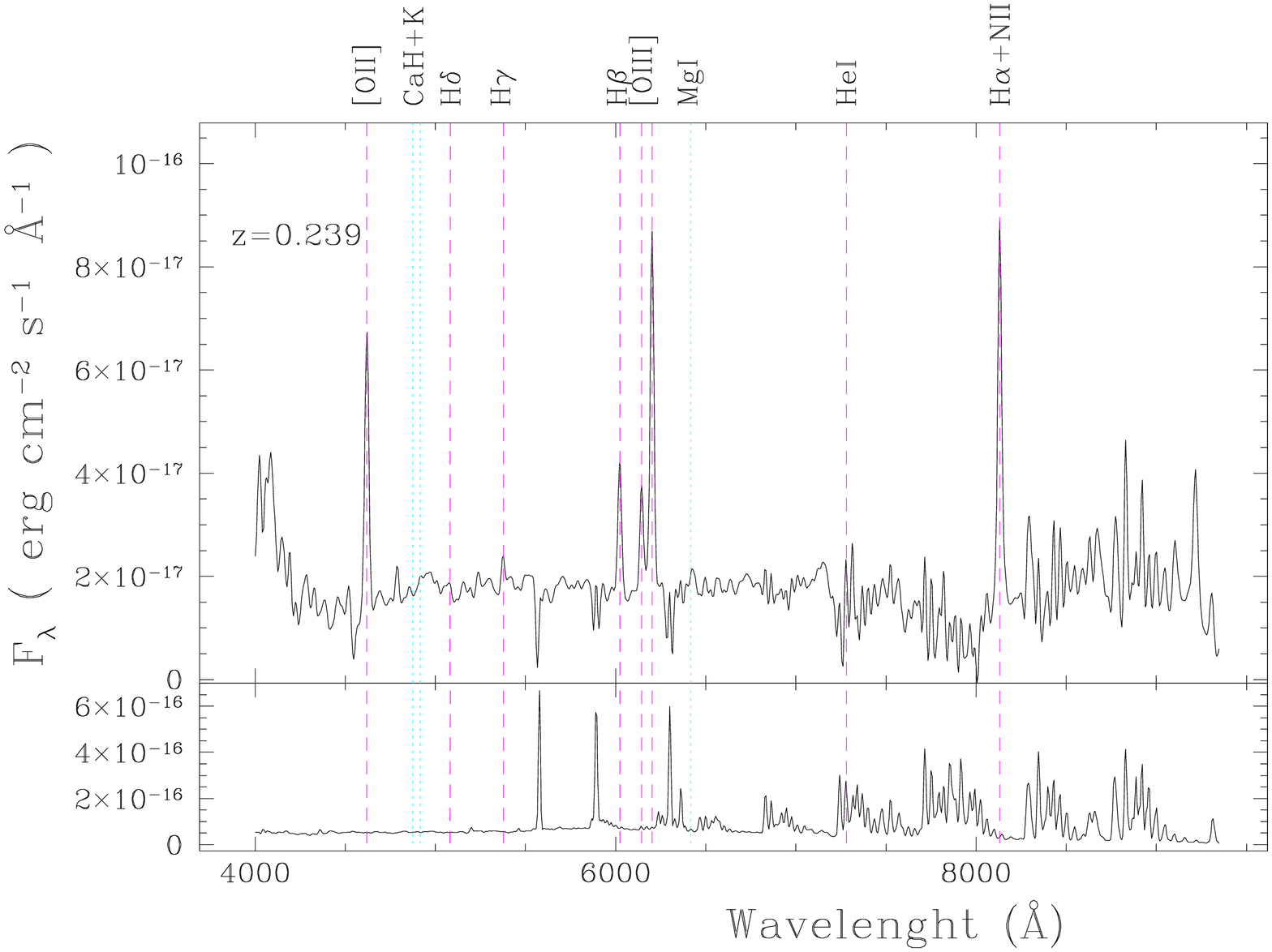}
\caption{
\label{fig:spectra}
Examples of three of the obtained spectra: from top to bottom, spectra of an 
early-type galaxy (entry number 22801), of a galaxy of intermediate type 
(entry number 42490), and of a late type galaxy (entry number 81046), at the
redshifts reported in each panel. Below 
each panel, the subtracted sky spectrum is shown. Dotted and dashed vertical 
lines show the positions of the absorption and emission lines, respectively,
whose identifications are given above the panels.
}
\end{figure}

\section{Observations}
\label{sec:obs}

Target galaxies were selected from the photometry in Zappacosta et al. (2002). 
We selected galaxies with \rr\ 
band magnitudes between 20 and 21.5. This magnitude range was chosen in 
order to maximize the detection efficiency around z$\sim$0.45. In fact, 
$L^*$ galaxies at this redshift are expected to
have $M^*($\rr$)=-21.5$ (Lin et al. 1999), corresponding to 
\rr=20.2 at z=0.45. The magnitude limit on the bright side is needed to exclude
nearby objects, given that galaxies significantly brighter than $L^*$ tend to 
be rare, while the limit of the faint side is needed not only to exclude
more distant galaxies, but also to limit the required exposure time.
The galaxy catalog comprises about 3000 galaxies within this magnitude range.
About 10\% of these were selected to be observed spectroscopically, 
trying both to maximize the number of galaxies near the X-ray peak 
and also to sample a large fraction of the region around it.
We used two different multi-objects spectrographs 
(see Table~\ref{tab:obs}) in order to maximize the number of observed 
objects both in the general field and on the X-ray peak. 

\subsection{William Herschel Telescope multifiber observations}

First, we used the AF2/WYFFOS spectrograph (Watson, 1995)
at the William Herschel Telescope (WHT).
This instrument is based on a bundle of 160 fibers that can be placed 
over an area of about 60\arcmin\ of diameter. 
The mounting of the fibers does not allow fibers to be placed
fibers closer than 25\arcsec\ from each other; as a consequence, we used this 
instrument to observe as many galaxies as possible over the whole field.
The large dimension of the fiber mounting also limited the number of 
fibers that can be used together to about 80 per pointing.
The fibers have a diameter of 90\mic, corresponding to 1.6\arcsec. 
For galaxies at z$\sim$0.4, the radius of this aperture (0.8\arcsec)
corresponds to $\sim$4.3 kpc,
to be compared with the half-light diameters of 1-5 kpc for 
elliptical galaxies 
and  1-4 kpc for late-type galaxies (e.g., Shen et al. 2003, see also 
Cresci et al. 2006).
For each pointing, 10 fibers are located in empty parts of the field 
and are used to sample the sky. Pointing and tracking are obtained by
using dedicated fiber bundles and are precise to within a few tenths of arcsec.
We used a low-resolution grating that provides a wavelength 
coverage between 4200 and 9300\AA\ with a resolution of R$\sim$600. 

Three fiber settings were used (see Table~\ref{tab:obs})
for a total of 237 galaxies. 
Target galaxies were split into three magnitude ranges in order to observe
all the galaxies with similar magnitude with a similar exposure time, and 
spend longer time on fainter objects. Exposure times were between 60 and 230 
min. For each image, a shorter (between 10 and 30 min) ``sky'' image was 
acquired, by displacing 
the telescope of a few arcsec with the same slit setting.  
A spectrophotometric standard star was observed through a few fibers 
to determine the average wavelength sensitivity of the instrument.

Special care was devoted to obtaining the best sky subtraction, which is usually
a critical step during data reduction of multi fiber data. 
We had two possible choices for the sky subtraction. 
First, we could use
the ``contemporary'' sky spectrum we obtained, together with the target 
spectra, through the 10 fibers
pointed in empty parts of the sky. The sky spectrum was taken in 
perfect temporal coincidence with the science spectra, but through 
different fibers. Second, we could use the spectra in the ``sky'' image 
taken after the observations. In the latter case we have an estimate 
of the sky taken with the same fibers as were used for the targets, 
but the temporal variation of the sky must be 
taken into account.
We decided to use the ``contemporary'' spectra because the temporal 
variation of the sky turned out to be dominant. 
Before subtraction, this sky spectrum was multiplied for a factor,
changing from fiber to fiber, obtained by minimizing the residuals 
on the bright sky lines. This step is needed to correct the spatial 
variations in the 
sky brightness and the variations in transparency of the fibers not 
corrected by the flat field.

The wavelength calibration, another critical step when measuring redshifts
with different instruments, was obtained by using images using a 
Neon lamp.
The uncertainty in the wavelength calibration, estimated by the RMS of the
residuals of the fit of the dispersion relation, is about 0.6\AA.

\subsection{Telescopio Nazionale Galileo (TNG) multislit observations}

The multifiber observations described above were complemented by other 
spectroscopic data obtained at the TNG with the DOLORES spectrograph.
This instrument uses multi-slit masks to observe up to $\sim$30 objects
in a field of about 7\arcmin$\times$9\arcmin. 
As a consequence, in this case we
have a lower multiplexing, but more objects can be observed in a small area,
in our case near the X-ray peak. 
Two masks were used, for a total number of 32 galaxies. 
We used slits with dimensions 1\arcsec.1$\times$20\arcsec\ and a 
low-resolution grating, providing a resolution of R$\sim650$ 
in the wavelength range between 4800 and 9500\AA.  For each
mask, an integration time of 150 min was used. The exposure time was split
into 2 or 3 images with a small telescope noddling along the slits. 
These images
were used to subtract the sky and to correct for cosmic rays. A second-order
sky subtraction was also obtained by spatially interpolating the long-slit spectra across the position of the targets. The wavelength
calibration was obtained by using the sky lines and an observation of an Argon
lamp. The resulting uncertainty is about 0.8\AA.

\subsection{Determining the redshifts}
\label{sec:redshift}

The redshift measure is based on the identification of emission lines, 
absorption lines, and spectral breaks.  When an emission line is present in 
a wavelength range without bright sky lines, it drives the redshift 
determination. 

From the average noise in the spectra, measured on part of the spectra
between the sky lines, we estimated that the typical 
emission line flux for detection is about $1.5\times10^{-16}$erg cm$^{-2}$
sec$^{-1}$ for both instruments. For the objects with
no detected emission line, we tried to estimate the redshift by comparing the
observed spectra with the templates in Mannucci et al. (2001) and
by using the spectral shape and the absorption lines. 
In many cases, the 4000\AA\ break, due
to the presence of evolved stars, is visible, together with the 
CaII $\lambda\lambda3934,3968$ doublet.
To each measured redshift we assign
a numeric quality flag, from 1 (certain, because based on unambiguous  
high-quality spectral features) to 2 (reliable, some of the spectral features
have low SNR or coincide with bright sky lines), 3 (uncertain), and 4
(tentative). 

Of the 269 observed galaxies, we determined 161 redshifts of quality 
classes 1 and 2 (101 of quality 1, including 3 QSOs, and 60 of quality 2) 
and 52 more redshifts of lower quality. One object turned out to be a 
Galactic star.
For the rest of this work, we only use 
the redshifts of classes 1 and 2 (certain and reliable).
The error in the redshift determination can be computed by considering
the uncertainties on both the wavelength calibration and the determination of
the spectral features. 
When an emission line is detected, the typical uncertainty in the 
determination of the line center is below 2--3\AA.
When the redshift relies on absorption lines,
the precision of the determination
on the position of the line depends on its equivalent width (EW), 
and is of the order of about 4--7\AA.
For all the redshifts of classes 1 and 2 the estimated error is 
below $\Delta z=0.002$ (corresponding to $\sim$8\AA\ for the \oii\ line at
z$\sim$0.4).

Galaxies have been classified in three classes according to their
spectral characteristics: early-type (pronounced 4000\AA\ break, 
CaII absorption lines with large EW, no emission lines), intermediate type
(presence of both 4000\AA\ break and weak emission lines), late type (strong
emission lines). Three objects show broad emission lines with 
a line width larger than 5000 Km/sec, and were classified as QSOs.
Fig.~\ref{fig:spectra} shows some examples of the obtained spectra, one
for each galaxy type.

The object catalog, together with the measured redshifts and the spectral
classifications, is reported in Table~2, published in the electronic 
version of this paper, and available at VizieR
(http://vizier.u-strasbg.fr/cgi-bin/VizieR). The columns of Table~2
report object identification, coordinates, \rr-band magnitude, used telescope,
measured spectroscopic redshift, redshift quality flag, and spectral
classification.

\begin{figure}     
\centering
\includegraphics[width=9cm]{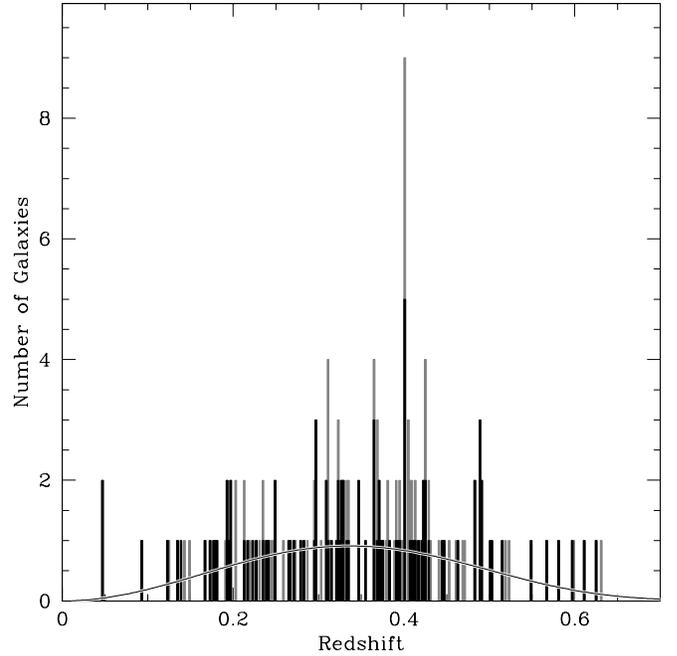}
\caption{
\label{fig:zhist}
Histogram of the redshift of quality classes 1 (98 galaxies, in black) 
and 2 (60 galaxies, in grey). 
The bin width is $\Delta z=0.002$, corresponding to the upper limit of the
redshift uncertainty.  The overplotted line shows the number of expected
galaxies per bin for a smooth universe  due to the magnitude selection of
the target galaxies}.
\end{figure}

\section{Redshift distribution}
\label{sec:zhist}

Figure~\ref{fig:zhist} shows the histogram of the measured redshifts with
quality classes 1 and 2, excluding the three quasars that have z$>$1.5. 
The distribution ranges between 0.05 and 0.63 
and peaks around z=0.40 as expected for the chosen limiting 
magnitudes of the sample (see Sect.~\ref{sec:obs}). 
The bin at z=0.401$\pm$0.001 contains
9 galaxies, a clear excess with respect to the nearby bins. 
The width of this bin corresponds to a velocity spread of $\Delta v=430$ km/sec
and, considering only the contribution from the Hubble flow, to a comoving 
depth of $\Delta r = 6.9$ Mpc.

To estimate the 
statistical significance of this result, we computed the expected number of 
galaxies from a universe with a flat distribution of galaxies 
using the luminosity function (LF) of Lin et al. (1999) and
taking into account the volume sampled at the different redshift, the
K-corrections (Fukugita, Shimasaku \& Ichikawa, 1995),
the photometric selection function, and the efficiency in measuring the
redshift. The result, normalized to the total number of objects, is shown in
Fig.~\ref{fig:zhist} and has a broad peak at z$\sim$0.35. 
In the z=0.401 bin, 0.84 galaxies are expected, and
this number does not critically depend on the hypothesis of the 
computation. In particular, it is not possible to significantly increase
the expected number of galaxies at z=0.401, 
since this value is already close to
the maximum of the distribution.
Using the Poisson distribution, we computed that there is 
less than $3\times10^{-7}$ probability to have 9 or more 
galaxies for a random fluctuation when 0.84 are expected. 
The probability that such an excess occurs in one of the 300 
considered bins
is, as a consequence, below $10^{-4}$.
An analogous result is obtained when considering only the 98
class 1 redshifts: 
in this case, 0.50 objects are expected and 5 are observed,
with a probability that the signal is due to a statistical fluctuation 
less than $4\times10^{-3}$.
A similar result is also obtained when restricting the analysis to the 43
galaxies within 6.5\arcmin\ from the X-ray peak (see next section).
In this case, 6 galaxies are found at z=0.401, while 0.22 are expected.
Also in this case the statistical probability of a chance signal is below
$5\times10^{-5}$.

As a consequence, this overdensity of galaxies cannot be 
ascribed to a random fluctuation in the number of galaxies but, 
on the contrary,
is the indication of a definite overdensity of galaxies
at z=0.401. Instead, all the other peaks in the redshift 
distributions are consistent with random fluctuations.

If a large-scale structure is present at this redshift, the galaxies in it
could have a velocity dispersion larger than the used bin width of $\Delta
z$=0.002
or could be dispersed across a larger region. 
In particular, this is true if the radial dimension of the detected
WHIM filament is larger than the dimensions on the plane of the sky. This is
likely to be the case, because the WHIM gas is optically thin to its 
own X-ray emission and,
therefore, the apparent surface brightness is actually given by the total
projected brightness. For this reason, filaments that ``point'' in our
direction could be easier to be detected. The dispersion of the
galaxy radial velocity across this region could be large, either because 
galaxies still follow the Hubble flow with different velocities or because
they have different bulk motions in different points of the filament.
Therefore we check if an overdensity is also present 
with a larger binning. When using a 5 times larger bin ($\Delta z=0.01$),
we find that the bin at z=0.405$\pm$0.005 contains 17 galaxies, to be
compared with 4.2 expected galaxies. In this case as well the probability of a
random coincidence is negligible. The overdensity is also present when using
a bin with $\Delta z=0.02$: in this case 23 galaxies have 
0.39$<$z$<$0.41, to be compared with the expected number of 8.7, with a 
probability of less than $5\times10^{-3}$. 

We checked that this overdensity of galaxies at z=0.40 is not due to an
observational bias, such as a particular configuration of the main
emission lines across the sky spectrum.
At this redshift, the main emission lines (\oii, \hb\, \oiii, and \ha) 
fall at 5222, 6810, 7015, and 9193\AA, respectively, and there is nothing 
at these wavelength that could justify a sharp increase in the redshift
determination efficiency.
As a conclusion, the large number of galaxies at z=0.401 is due to the presence
of a real large-scale structure of galaxies at that redshift.

It should also be noted that 
Postman et al. (2002) detected as many as
6 galaxy clusters with spectroscopic redshift
between 0.38 and 0.40 
within 2 deg (about 40 Mpc) from the X-ray peak.
This is 1/3 of their total sample, as in this area 
they detected and spectroscopically confirmed 18 galaxy clusters.
As a consequence,
the overdensity of galaxies at z=0.40 and the WHIM filament
could be linked to a complex cosmological structure present 
on a much larger scale and traced by the observed clusters. 

\begin{figure}     
\centering
\includegraphics[width=9cm]{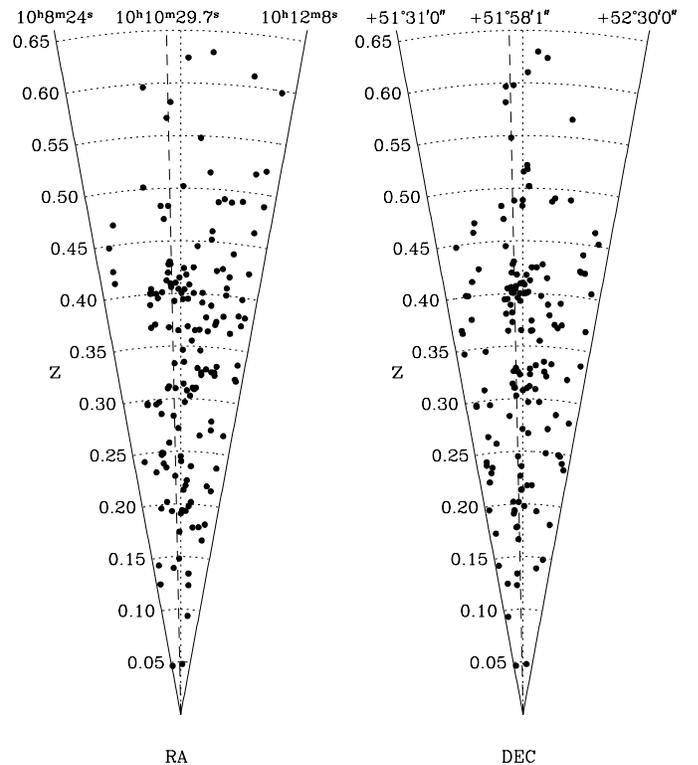}
\caption{
\label{fig:cone}
Cone diagram of the position on the targets with measured redshift 
(quality classes 1 and 2). The redshift (vertical axis) is plotted vs. the
R.A. (left cone) and the Declination (right cone). The dashed line shows the
position of the X-ray peak, and the dotted line is the center of the field. 
The labels on the tops show the coordinates of the limits of the field and of
the X-ray peak.
}
\end{figure}

\begin{figure}     
\centering
\includegraphics[width=8.8 cm]{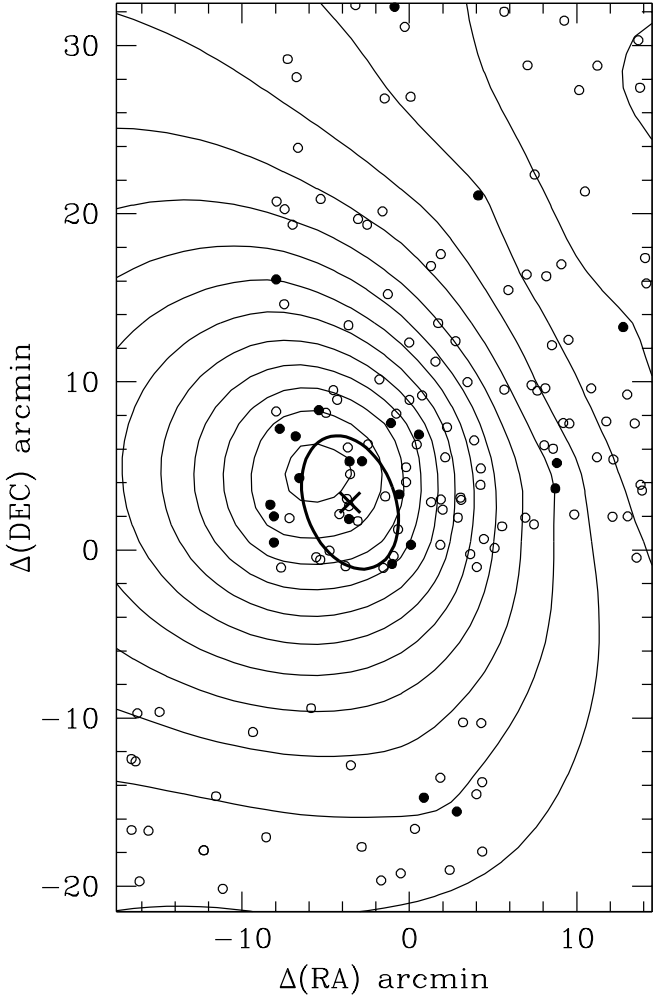}
\caption{
\label{fig:chance}
The contour plot showing the average distance of the 23 galaxies at 
0.39$\le$z$\le$0.4
from each point of the field normalized by the average distance, from that
point, of all the target galaxies. This value is a measure of the degree of
clustering around any given point of the field, taking into account the 
inhomogeneous distribution of the target galaxies.
The contour values goes from 0.65 to 1.04
with steps of 0.03. The dots show the position of the target galaxies, with
black dots showing the galaxies at z$\sim$0.4. The black cross is the position
of the X-ray peak, with the ellipse showing its half-maximum extension.
The distribution of the normalized average distance has a well-defined minimum
(0.64) consistent with the position of the X-ray peak.
}
\end{figure}

\begin{figure}     
\centering
\includegraphics[width=9cm]{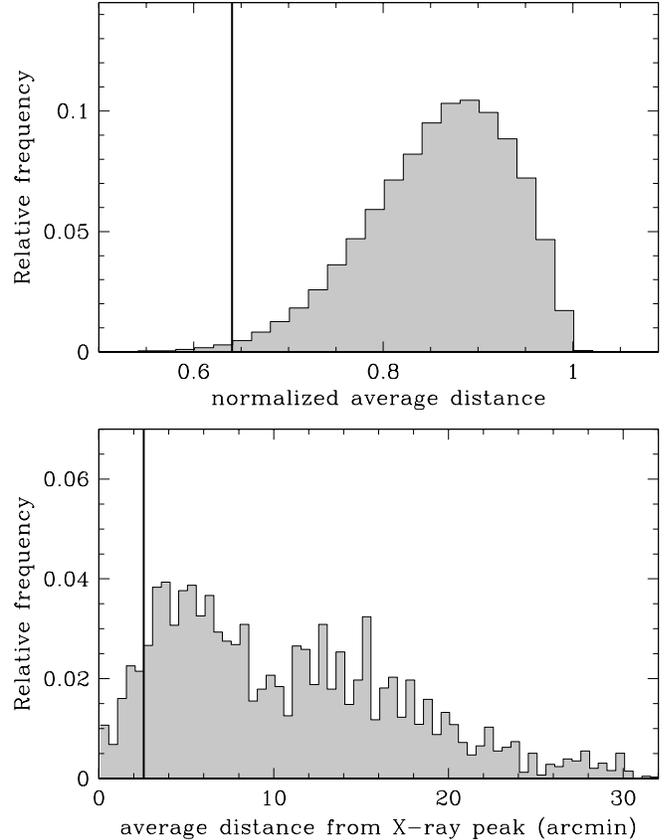}
\caption{
\label{fig:plotrand}
{\em Upper panel:} 
the distribution of the minimum distance for 100000 random samples
of 23 galaxies (grey histogram, see also Fig.~\ref{fig:chance}).
Only 0.7\% of the random samples shows a value equal to or smaller than 
that for the 23 galaxies at 0.39$\le$z$\le$0.41, shown by the
the thick vertical line.
{\em Lower panel:} distance from the X-ray peak of the position of the 
centers of the 10\% of the random samples showing the highest degree of
clustering, compared to the value for the 23 galaxies at z$\sim$0.4 (thick
line).
}
\end{figure}

\begin{figure}     
\centering
\includegraphics[width=9cm]{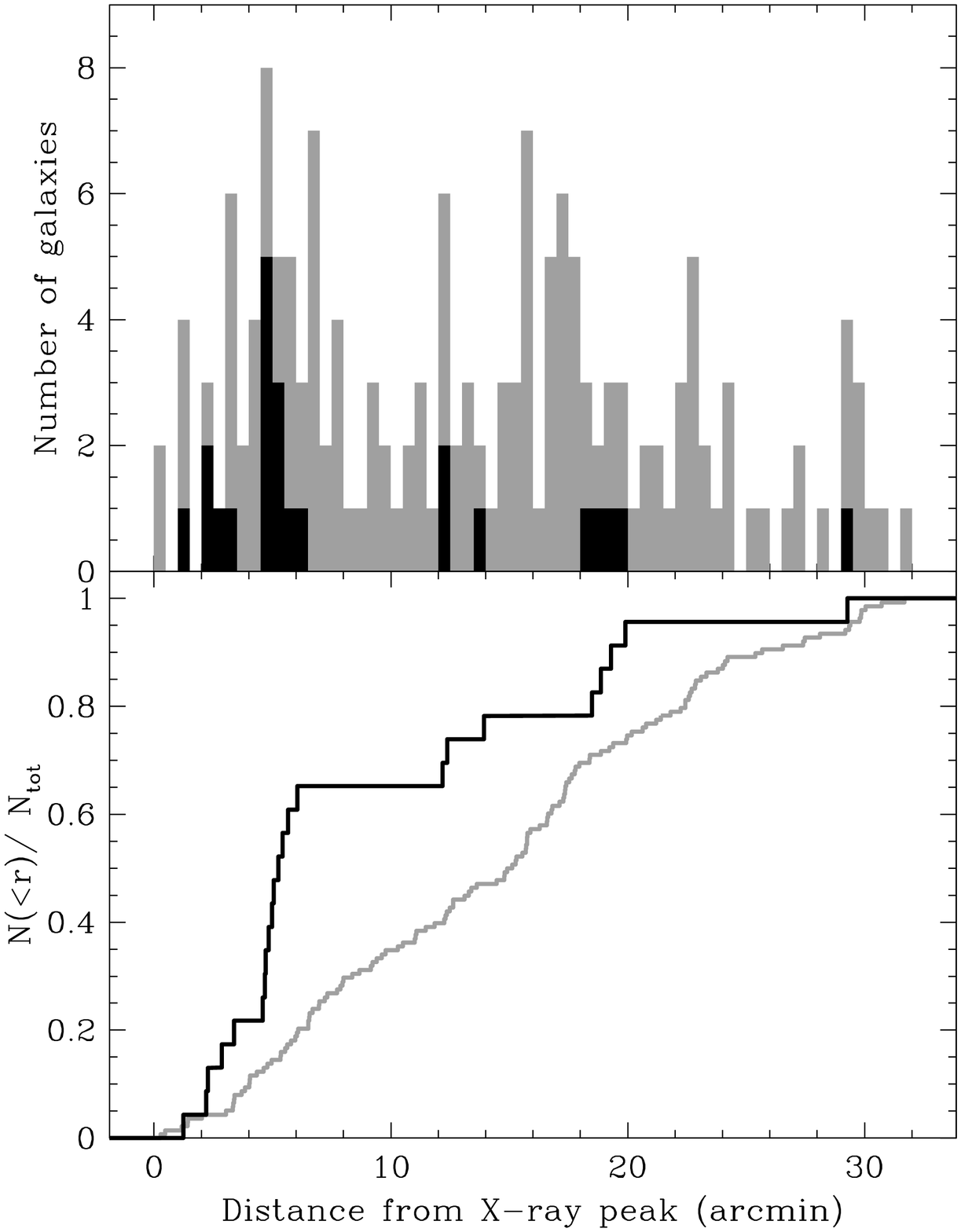}
\caption{
\label{fig:dist}
{\em Upper Panel}: distribution of the distances from the X-ray peak of
the 23 galaxies having 0.39$<$z$<$0.41 (in black) and all
the 158 galaxies with a measured redshift (in grey). 
Only galaxies with a quality class 1 or 2 are used.
{\em Lower Panel}: comparison of the normalized cumulative distributions of
distances (lines color-coded as above), showing that the galaxies belonging to
the redshift peak are much more concentrated toward the X-ray peak than the
general sample.
}
\end{figure}

\section{The spatial distribution of the galaxies at z=0.40}
\label{sec:spatial}

The presence of peaks in the redshift distribution of objects is a very common
feature of many redshift surveys (see, for example, Steidel et al.
2000) and is known to be due to the large-scale structure of the universe. In
this section we intend to study the spatial 
distribution of the galaxies belonging to the redshift peak
to secure the spatial (and therefore physical)
association between the X-ray peak and the galaxy
overdensity. Zappacosta et al. (2002) have already shown that there is a
good spatial coincidence between the X-ray peak and the galaxy overdensity
as revealed both by the total number counts and the photometric redshifts
in the range $0.3<z<0.6$. (see their Fig.~6). 
In the following, we demonstrate that the galaxies with a spectroscopic
redshift z$\sim$0.4 have a distribution centered on the X-ray peak
and, as a consequence, are spatially associated with it.

As seen in Fig.~\ref{fig:targets},
the galaxies in the redshift peak are more
spatially clustered around the X-ray peak than the rest of the galaxies,
and in the previous section we have already noted 
that 6 out of the 9 galaxies at z=0.401 are found within 6.5\arcmin\ from the
X-ray peak.
The presence of clustering in both the redshift and projected positions
can be better seen in Fig.~\ref{fig:cone},
where each one of the two spatial coordinates R.A. and
Dec. are plotted as a function of the redshift. 

In the following we intend to statistically verify that the
23 galaxies at 0.39$\le$z$\le$0.41 are spatially clustered and that their
cluster center is compatible with the X-ray peak. The computation
must take the spatial distribution of all the targets galaxies into account,
that are not homogeneously distributed across the field.

If a group of galaxies are spatially clustered, a point of 
the field exists with the properties of a ``center of the cluster''; 
i.e., the average distance
of the galaxies of the selected group from this point 
must be significantly smaller
that the average distance of all the galaxies of the full sample.
For this reason
we computed the average distance of the 23 galaxies at z$\sim$0.4
from any given point of the field and normalized
the result to the average distance of all the 158 target galaxies 
with a measured redshift (classes 1 and 2) from this
point\footnote{As a comparison sample,
we do not consider the full target sample
because the efficiency of determination of the redshift in the WHT multifiber
data decreases sharply with the distance from the field center. This is due to
the fact that the WHT PSF degrades with the radial distance and the fraction
of light impinging in the 1.6\arcsec fiber reduces with the distance from the
optical axis, which falls near the X-ray peak. As a consequence, the comparison
with the full target sample would reveal a false concentration of the galaxies 
near the X-ray peak.}.
This process compares the distribution of the selected galaxies to that
of the total sample and therefore takes the distribution of the
target galaxies into account.
If the considered group of galaxies is not spatially clustered, 
the average distance from all the points in the field will be similar to that
of all the target galaxies; i.e., the ratio between these two distances
will be compatible with 1
for all the points in the field.
If, in contrast, the selected galaxies are spatially clustered, 
a point of the field will exist for which 
this ratio is considerably lower than 1, and this point can be considered to
be the ``center'' of the cluster. Figure~\ref{fig:chance} illustrates
this computation for the 23 galaxies at z$\sim$0.4. The distribution of the
normalized average distance shows (1) the presence of a well-defined 
minimum of 0.64, meaning that the galaxies at z=0.4 are more clustered 
than all the target galaxies, and (2) the position of this center
turns out to be consistent with the X-ray peak.

To verify that these results are not due to the target distribution, we have
extracted 100,000 random samples of 23 galaxies from the target list and 
repeated the computation in Fig.~\ref{fig:chance}
to see (1) how many random groups of galaxies appear to be as clustered 
as the galaxies at z$\sim$0.4, and (2) what the spatial
distribution of the centers is for the random groups showing a high
degree of clustering.
The results are shown in
figure~\ref{fig:plotrand}. In the upper panel, the distribution of the 
values of the minima is compared to the value (0.64) obtained for the 
galaxies at z=0.4. Only 0.7\% of the random sample have normalized 
average distances below
this value, demonstrating that (1) the galaxies at z$\sim$0.4 are not a
random sample and that (2) these galaxies show a strong clustering that
cannot be attribute to the target distribution. Note that 
{\em this result is obtained without any assumption on the position of the
center}, i.e., we are not using the position of the X-ray peak in any way.
In the lower panel of Fig.~\ref{fig:plotrand} we concentrate
on the random samples showing a significant clustering (values of the minimum
ratio below 0.75, corresponding to the 10\% random galaxy groups
showing the highest clustering). For these groups we show the distance 
of the derived center from the X-ray peak. Only 7\% of them 
have centers closer to the X-ray peak than the galaxies at z$\sim$0.4, 
demonstrating that the target distribution allows the potential detection of
clustering over the whole field. As a consequence, the clustering of
the galaxies at z$\sim$0.4 around the X-ray peak is due neither to a
statistical fluctuation nor to the distribution of the target galaxies.

\smallskip

A simpler and more direct way to statistically verify that galaxies 
at z$\sim$0.40 are spatially associated with
the X-ray peak consists in computing their average distance from this point
and comparing this with the average distance of all the target galaxies.
Also, this method takes the irregular distribution of the targets into account,
while only clustering around the X-ray peak is considered.

The histogram of the distances from the X-ray peak of the 23 galaxies 
at z$\sim$0.40 and of the target galaxies
is shown in upper panel of Fig.~\ref{fig:dist}.
To compare the two samples of galaxies, we used two statistical tests.
First, we used the Kolmogorov-Smirnov test to compare the two cumulative
distributions of distances from the X-ray peak, as shown in the lower panel
of Fig.~\ref{fig:dist}. From this test we derived that
there is only a 0.03\% probablity that 
galaxies at z$\sim$0.4 have the same spatial distribution of the total sample.
This demonstrates a spatial association between
galaxies at z$\sim$0.4 and X-ray peak with a confidence level above 99.97\%.

This is also confirmed by a Montecarlo simulation. We computed the
average distance from the peak of the 23 galaxies at z$\sim$0.4, 
finding 9.9\arcmin. 
Then, we extracted $10^5$ random samples of 23 objects 
from the galaxies with redshift and computed the average distance from 
the X-ray peak of these galaxies. 
The distance distribution of the random samples
has a Gaussian shape centered at 13.7\arcmin\ and with $\sigma$=1.7\arcmin. 
As a result, only
0.2\% of the random samples have an average distance from the X-ray peak equal
to or below that of the 23 galaxies at z$\sim$0.4. 
This demonstrates that this is not a
random sample with a confidence level of 99.8\%, in good agreement with the
Kolmogorov-Smirnov test.

Summarizing, we have demonstrated the existence of a galaxy overdensity
at z=0.40 and its spatial coincidence with the soft X-ray emission. 
The possibility of the X-ray emission being unrelated to the galaxy 
overdensity
cannot be completely ruled out, but the results of the statistical
tests make a chance association very unlikely.

Figure~\ref{fig:dist} shows that most of the galaxies with z$\sim0.40$
fall within 6.5\arcmin\ (about 2 Mpc at z=0.4) from the X-ray peak. 
It should be noted that the spatial distribution of our targets
is not homogeneous and, in particular, no galaxies were observed 
more than 5\arcmin\ to the east and more than 4\arcmin\ to the south 
of the X-ray peak. As a consequence, the galaxy structure 
could be much more extended in
these directions, and the observed dimensions of about 2 Mpc
must be considered a lower limit to the real projected dimensions.

\smallskip

Zappacosta et al. (2002) have shown that the temperature of the gas
emitting the X-ray peak is too low to be due to a cluster. As a consequence,
the region must have only a limited overdensity, and we expect that the
galaxies have characteristics similar to the field. 
We cannot estimate the surface density of galaxies at  0.39$<$z$<$0.41,
since our spectroscopic sample covers only about 10\% of the galaxies
in the selected magnitude range.
A sensitive indicator of the environment is the spectral classification 
of the galaxies:
in clusters, the fraction of early-type galaxies is roughly
0.6-0.9 (Treu et al. 2003), while in the field it is about 0.4  
(Smith et al. 2005). 
The galaxy classification is present for just 45\% of the observed
galaxies and, as a consequence, can only provide limited information.
Nevertheless, the results do not contradict the expectations,
as only 3 out of the 23 galaxies with 0.39$<$z$<$0.41 are classified as early
type galaxies. Even assuming that all the 9 unclassified galaxies
are early type, the fraction of such galaxies remains below 48\%.
Limiting this analysis to the galaxies within 6.5\arcmin\ of the X-ray
peak, this fraction is limited to below 8/15=53\%.

\begin{figure}     
\centering
\includegraphics[width=9cm]{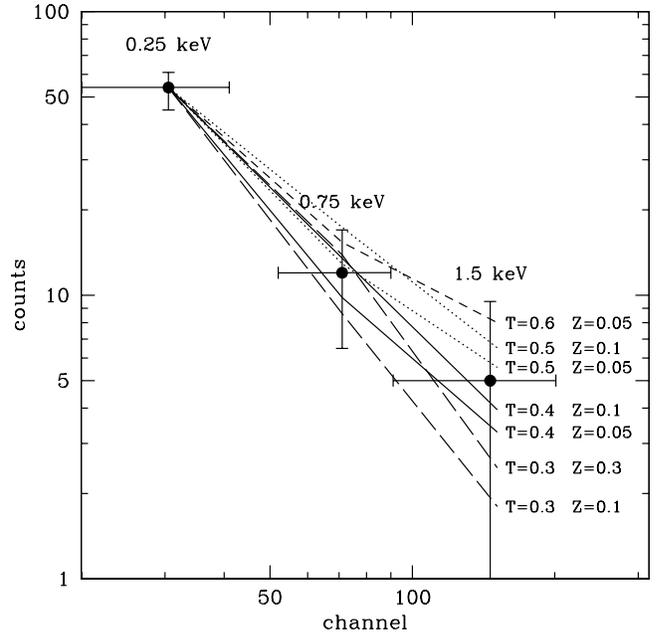}
\caption{
\label{fig:sed}
The observed values of the X-ray flux in the three ROSAT channels (dots),
de-absorbed for Galactic hydrogen,
are compared with the expectations from a hot plasma. 
Several possible fits are present, and each line is labeled
with the value of the corresponding temperature T (keV) and abundance Z 
(relative to solar).
}
\end{figure}

\section{Discussion and conclusions}

Zappacosta et al. (2005b) used the observed values of the 
WHIM temperature to constrain its thermal history and compare it with the 
results of the theoretical models. 
Such a comparison is crucial for testing the cosmological
models in one of their most important predictions. Unfortunately, the
uncertainties in the observations are still very large and the comparison 
is not yet able to put strong constraints to the model. Moreover, the detection
could be dominated by selection effects, in that only the densest and hottest 
regions are detected, as these provide the brightest X-ray emission.

Our determination of the spectroscopic redshift of the galaxy overdensity
associated to the WHIM structure removes one of the main uncertainties  
in determining of the temperature. As a consequence, we can now 
refine the temperature determination in Zappacosta et al. (2002) 
that was based on a very uncertain photometric redshift. 
In particular, we can now study the dependence of the 
best-fitting temperature with the metallicity that is assumed for the IGM. 
This quantity is very poorly constrained by the models 
(e.g., Viel et al. 2005).
Zappacosta et al. (2002,2005) estimated a value of $\sim$0.3 keV 
by assuming an abundance of 0.3 solar, similar to what is
measured in the external part of the galaxy clusters (De Grandi et al. 2004).
In fact, clusters of galaxies poorly
represent the typical physical conditions of the WHIM
environment (i.e. temperature, gas, and galaxy density). The outskirts
of groups of galaxies can give a fairer approximation of the
typical conditions in WHIM environments.
Recently Buote et al. (2004)
made a measurement of the metallicity in the distant regions
(i.e. 20-40\% of the virial radius) of the group of galaxy
NGC~5044. They measured a metallicity of about 0.1 solar (assuming
Anders \& Grevesse 1989 solar abundances), significantly lower than the
typical values measured in cluster outskirts.
In Fig. \ref{fig:sed} we present the results of the fitting of the X-ray
SED using an APEC plasma code 
by allowing metallicity to vary between 0.05 and 0.3 solar.
A strong degeneracy between metallicity and temperature is obtained, i.e,
a decrease by a factor of six in the assumed metallicity 
corresponds to an increase of a factor of two in the 
best-fitting temperature.
As a consequence, temperatures between 0.3 and 0.6 keV
can reproduce the observed SED equally well.

We emphasize that several other uncertainties remain that could
significantly change the temperature estimate. The main one is
due to the uncertainties in the background subtraction, 
as described in Zappacosta et al. (2002), which could alter the 
soft X-ray flux up to a factor of two. Even if the existence of this
X-ray excess is robust, its observed spectral shape can be significantly
affected by the assumed level of the background.
Despite these large uncertainties, it is evident that these values 
of temperature and metallicity are in good agreement with the 
expectations for the WHIM (Viel et al. 2005; Dav\'e et al. 2001), while
other possibilities have been already discarded by Zappacosta et al. (2002).

In conclusion, a spectroscopic observation of the galaxies projected 
near a peak of diffuse soft X-ray emission has managed to confirm the
presence of the galaxy overdensity associated with this structure and,
as a consequence, its extragalactic nature. The temperature 
of the X-ray emitting gas can be constrained to between 0.3 and 0.6 keV,
despite the large uncertainties in the data points and the degeneracy with
the metallicity of the IGM medium.
The spatial distribution of the galaxies and the temperature of the gas 
demonstrate that the X-ray emission is due to WHIM at z=0.40
having a temperature of 0.3--0.6 keV.

\begin{acknowledgements}
We are grateful to the TNG staff and to R. Corradi for assistance during 
observations, and to L. Magrini for useful discussion about data reduction.
\end{acknowledgements}

{}

\end{document}